\documentclass{article}

\usepackage{arxiv}
\usepackage{amsmath, amssymb}
\usepackage{siunitx}  %for units, e.g.\si{\volt},\si{\ampere}
\usepackage{subcaption}
\usepackage{caption}   % for \captionof, to avoid float figures in the appendix to overflow

\usepackage[utf8]{inputenc} % allow utf-8 input
\usepackage[T1]{fontenc}    % use 8-bit T1 fonts
\usepackage{hyperref}       % hyperlinks
\usepackage{url}            % simple URL typesetting
\usepackage{booktabs}       % professional-quality tables
\usepackage{amsfonts}       % blackboard math symbols
\usepackage{nicefrac}       % compact symbols for 1/2, etc.
\usepackage{microtype}      % microtypography
\usepackage{lipsum}		% Can be removed after putting your text content
\usepackage{graphicx}
\usepackage{natbib}
\usepackage{doi}

\title{A Quantifiable Information-Processing Hierarchy Provides a Necessary Condition for Detecting Agency}

\author{
Brett J. Kagan\thanks{These authors contributed equally.} \thanks{Corresponding author.} \\
Cortical Labs, \\
Melbourne, VIC 3000, Australia. \\
Department of Biochemistry and Pharmacology, \\
The University of Melbourne, \\
Parkville, VIC, Australia. \\
\texttt{brett@corticallabs.com}
\And
Valentina Baccetti\footnotemark[1] \\
Department of Physics, School of Science, \\
RMIT University, \\
Melbourne, VIC 3001, Australia. \\
RMIT Applied Quantum Technologies Centre, \\
RMIT University, \\
Melbourne, VIC 3001, Australia.
\And
Brian D. Earp \\
Centre for Biomedical Ethics, \\
Yong Loo Lin School of Medicine, \\
National University of Singapore, \\
Singapore 117597. \\
\And
J. Lomax Boyd \\
Berman Institute of Bioethics, \\
Johns Hopkins University, \\
Baltimore, MD, USA. \\
\And
Julian Savulescu \\
Centre for Biomedical Ethics,\\ 
\phantom{\texttt{email}} Yong Loo Lin School of Medicine, \phantom{\texttt{email}} \\
National University of Singapore, \\
Singapore 117597. \\
Uehiro Oxford Institute, \\
University of Oxford, \\
Oxford, OX1 1PT, England.
\And
Adeel Razi \\
Turner Institute for Brain and Mental Health, \\
Monash University, Clayton, Australia. \\
Wellcome Centre for Human Neuroimaging, \\
University College London, \\
London, UK.
}

\date{November 17, 2025}

\renewcommand{\shorttitle}{A QUANTIFIABLE INFORMATION-PROCESSING HIERARCHY}

\hypersetup{
pdftitle={A template for the arxiv style},
pdfsubject={q-bio.NC, q-bio.QM},
pdfauthor={David S.~Hippocampus, Elias D.~Striatum},
pdfkeywords={First keyword, Second keyword, More},
}

\begin{document}
\maketitle

\begin{abstract}
As intelligent systems are developed across diverse substrates - from machine learning models and neuromorphic hardware to in vitro neural cultures -understanding what gives a system agency has become increasingly important. Existing definitions, however, tend to rely on top-down descriptions that are difficult to quantify. We propose a bottom-up framework grounded in a system’s information-processing order: the extent to which its transformation of input evolves over time. We identify three orders of information processing. Class I systems are reactive and memoryless, mapping inputs directly to outputs. Class II systems incorporate internal states that provide memory but follow fixed transformation rules. Class III systems are adaptive—their transformation rules themselves change as a function of prior activity. While not sufficient on their own, these dynamics represent necessary informational conditions for genuine agency. This hierarchy offers a measurable, substrate-independent way to identify the informational precursors of agency. We illustrate the framework with neurophysiological and computational examples, including thermostats and receptor-like memristors, and discuss its implications for the ethical and functional evaluation of systems that may exhibit agency.
\end{abstract}

% keywords can be removed

\section{Background \& Scope}
Interest and investment in developing autonomous and intelligent systems have increased rapidly in recent years \cite{wong_interrogating_2025, betz_why_2025}. These advances have generated uncertainty and even disagreement about how to define key terms describing such systems \cite{kagan_scientific_2023, wong_interrogating_2025}. Among the most contested is the concept of \textit{agency}. Although many related terms have been flagged as requiring explicit definition \cite{kagan_toward_2024}, we focus here on agency as it applies to artificial and natural systems alike. In this paper, we do not explicitly endorse any single definition of agency. Instead, we put forward an empirically testable \textit{necessary} (if not sufficient) condition for an ostensibly intelligent system to possess agency on a range of plausible definitions that have been offered. In contemporary AI research, development, and marketing, agency has become a common yet inconsistently applied descriptor \cite{dung_understanding_2025, floridi_ai_2025, van_lier_introducing_2023}. To resolve ambiguity,  we propose to adopt a substrate-independent perspective, according to which two systems that exhibit the same informational properties should be described using the same conceptual vocabulary, regardless of their physical composition \cite{dattathrani_concept_2023}. This commitment concerns only descriptive equivalence; it does not imply moral equivalence. Differences in moral status may depend on which—and how many—properties are instantiated \cite{kagan_embodied_2024, sebastian_first-person_2021}. Ethical implications are noted briefly below, but a full analysis of how combinations of features map onto moral standing lies beyond the present scope. Our aim is instead to provide a formal, quantifiable framework for identifying where agency may arise, based on the intrinsic informational dynamics of a system.

We propose that agency definitions can be constrained by a system’s information-processing order — that is, the extent to which its internal transformations of input depend on its own prior states and outputs.  In short, we propose that for a system to have agency, the information processing architecture of that system is able to effect a qualitative, non-invertible change to the received information where this change itself is adaptable over time. We term this a Class III system and contrast it to proposed Class I and Class II systems. This measure does not offer a complete definition of agency but rather a necessary informational condition for its potential emergence. Consistent with the broader literature on situated and embodied systems \cite{dodig-crnkovic_systematic_2024, barandiaran_defining_2009, pickering_what_2024}, we consider potential agents as entities embedded within a larger informational environment. Following prior work \cite{sultan_bridging_2022}, we also distinguish agency from behaviour: agency concerns the capacity to act, whereas behaviour refers to the manifestation of that capacity. Building on these premises, we adopt a bottom-up approach that examines how a system’s internal dynamics respond to and integrate information from its environment. Therefore, differences in agency can be constrained based on differences in information-processing order, ranging from purely reactive mappings to adaptive, self-modifying dynamics that underpin goal-directed or autonomous behaviour.

The paper proceeds as follows. In Section 2, we outline differing approaches to determining agency and justify why a system theory approach can resolve ambiguity in the current approaches. In Section 3, we introduce the framework for order of information processing and provide minimal mathematical definitions. In Section 4, we provide case examples for each type of information processing that we propose, and show in simulated models how input is changed by the system. In Section 5, we then show how the memory and adaptivity of these different systems can be directly compared to each other over time. Finally, in Section 6, we explore the applicability of the proposed information-processing order frameworks in key areas before offering a conclusion.

\section{Differing approaches to determine agency}

Agency has been defined in various ways. Generally, a system is thought to have agency insofar as it has one or more of the following features: goal-directedness \cite{dung_understanding_2025}, a first-person perspective or sense of ownership over decisions  \cite{das_agency_2025}, independence to act to achieve goals  \cite{dodig-crnkovic_systematic_2024}, the ability to model its own activity and/or activity in its environment \cite{barzegar_minimalist_nodate}, or the capacity to drive intentional (where “intentional” means anything the agent desires to do) actions as per internal mental models \cite{swanepoel_does_2021, swanepoel_artificial_2024}. When considering the capacities that may define agency, these can each be considered at different levels of abstraction, taking into account features such as: interactivity, autonomy, and adaptability in addition to any of aforementioned features \cite{floridi_ai_2025, floridi_morality_2004}. A key limitation of existing accounts is that they define agency from the top down—by reference to a system’s capacity for certain kinds of behaviour—rather than from the bottom up, in terms of the mechanisms that generate that capacity. This creates a circular problem: if agency is inferred only from outward behaviour, then behaviour itself becomes the de facto marker of agency. Yet behaviour can easily be misleading about what is happening internally. As a result, systems may be described as “agents” based on their external performance alone, even when the underlying processes differ entirely from those that would constitute genuine agency. This risks conflating behavioural mimicry with true agency and, in turn, invites misplaced ethical attributions of moral agency, responsibility \cite{veliz_moral_2021, nyholm_ethics_2021}, or moral patiency \cite{shevlin_how_2021}. This is further complicated by fuzzy boundaries between a system and the external environment \cite{aguilera_agency_2018}, especially for nested systems; where systems of information processing operate inside larger systems of information processing. These nested systems constitute a broad class of systems that would include all biological, and many artificial, systems. Ultimately, without considering the underlying mechanisms which give rise to these capacities, it is difficult to determine in practice whether a given system has agency, and if so, of what type, or to what extent. 

\subsection{Why behaviour alone is inadequate for determining agency}

Assessing agency requires more than observing behaviour; it also involves understanding a system’s potential capacity to act under different conditions. To illustrate this reasoning, consider the case of counterfactuals—imagined alternative outcomes \cite{byrne_counterfactual_2016, dwyer_fundamental_2021}. Behaviour alone can make it hard to judge a system’s underlying agency, since external factors may prevent it from acting as it otherwise could. For example, in a game of poker, the best player might still lose because of bad luck. Yet we can imagine a counterfactual world where, with different cards, that same player would likely have won—because their decision-making was in fact superior. This shows that an agentic system may possess the capacity to achieve certain outcomes even if, in a given environment, those outcomes are not realized. Extending this idea, even systems we regard as clearly agentic — such as the human brain — depend on external conditions for expression. The brain must be embodied in a functioning body, which acts as its interface with the world. The brain’s capacity for agency is not diminished simply because the body, under certain constraints, cannot carry out the brain’s intended actions.

Here is another angle on the matter.  Agents are often described as maximizing some form of utility or reward. Whilst this is generally true, systems with the same outward behaviour can differ greatly in their internal mechanisms \cite{ramirez-ruiz_complex_2024}. For instance, agents guided by the maximum occupancy principle—which drives them to explore and occupy a wide range of action–state paths—can perform as well as reward-maximizing agents but show far more diverse behaviour \cite{ramirez-ruiz_complex_2024}. Such agents could be said to have greater agency, since their actions arise from internal drivers rather than rigid external goals, giving them more flexibility and choice. This idea aligns with findings that humans report a stronger sense of agency when they can compare and select among alternative actions \cite{kulakova_i_2017}. In general, systems governed purely by external rules exhibit less genuine agency than those whose internal dynamics allow them to choose among multiple possible courses of action \cite{hendrickx_agentially_2023, ramirez-ruiz_complex_2024, kishore_chakrabarty_causal_2025}. Complex systems, including those of human agents, can also exhibit hierarchically nested goals that, collectively, are regarded as necessary for well-being \cite{miller_predictive_2022}. 
 
Finally, consider a classic example often used in discussions of agency: the thermostat. It exists within an information environment and adjusts its output in response to changes in that environment. Moreover, moderns thermostats implement Proportional-Integral-Derivative algorithms to predict and achieve thermostatic goals. Does this make it an agent? Some capacity-based accounts would say yes, at least for sufficiently complex thermostats \cite{floridi_ai_2025}. Yet, intuitively, most people would not want to conclude that a thermostat is a true agent. Unlike animals or learning systems, a thermostat lacks internal dynamics that let it evaluate or act on alternative possibilities with intentionality—it merely reacts. Without resorting to anthropomorphism, this shows the limits of defining agency purely in terms of behavioural capacity. \footnote{Some have suggested that agency depends on the observer’s chosen frame of reference \cite{abel_agency_2025}, but this view makes agency a property of perspective rather than of the system itself. On such grounds, agency could no longer serve as an objective feature of a system.}  

\subsection{Supplementing with a system theory approach can resolve ambiguity}

An alternative approach is to take a system theory approach to agency, considering how agency (as understood on a range of plausible definitions) may emerge based on features of the system \cite{miehling_agentic_2025}. This approach seeks to show how a complex whole emerges from an interaction of constituent parts \cite{astrom_feedback_2021, hofkirchner_general_2011}. That is, it examines how each component functions on its own and how it interacts with others to produce the system’s overall behaviour. Unlike a gestalt view — which assumes that the whole is always greater than the sum of its parts — a systems theory perspective holds that complex or non-linear outcomes may emerge from the interaction of parts, but need not do so. To develop a systems theory account of agency, we must identify the fundamental feature shared by all systems that can display it. As noted above, such systems exist within a broader information environment and are affected by it. Any system with the potential for agency must be able to receive and process external information through some sensory mechanism that links the environment to its internal states. Crucially, systems with agency must handle this information in a qualitatively different way from those without it. We propose that there are three ordered classes of information processing within systems, and that genuine agency arises only in systems capable of the third and highest class, which encompasses all lower levels as well.

\section{Orders of information processing}

Information-processing 'orders' refer to different complexities in how information is processed by a system. The word 'order' is used to capture that each higher order also contains the complexity of the orders below it. Here, we proposed the idea of orders of information processing as a structural framework for distinguishing systems by how they mediate and transform information through the interplay of input, internal state, and output. While in theory there could be systems with unbounded orders of information processing, we propose that these can be simplified into three classes that also correspond to the first three successive orders of information-processing behaviour. 
\begin{itemize}
\item Class I systems are purely reactive, mapping inputs directly to outputs and are memoryless. \item Class II systems extend the function of Class I systems to incorporate internal states that allow their responses to depend on both current and past inputs, yet the way they combine or weight this information remains fixed over time. 
\item Class III systems further extend upon Class II systems to include adaptive mechanisms through which the parameters governing input–state coupling evolve in time as a function of the system’s own activity to allow self-modulation.
\end{itemize}
Class III systems may have multiple levels of adaption which can greatly increase the complexity. As the complexity of information processing increases within a Class III system, the information-processing dynamics may become progressively shaped by the system’s own history and configuration, giving rise to properties such as memory and adaptivity. Yet, for the purpose of establishing a necessary condition for a system to be considered agentic, these additional mechanisms are not required. Finally, we also acknowledge that higher class systems could also exist if qualitatively different transformations of information occur, yet considering these dynamics is beyond the scope of this paper and current focus.  

The concept of information-processing classes provides a structural basis for distinguishing systems according to how they handle and transform information. Rather than focusing on learning or performance, this approach examines the intrinsic relationships between input, internal state, and output. As the complexity increases, these relationships become progressively more dependent on prior inputs and internal configurations, leading to the emergence of memory and adaptivity. This hierarchy offers a principled way to identify the necessary informational precursors of agency. It is also necessary to view the different classes of a system in the hierarchical framework, as Class III systems will almost certainly contain Class I and Class II elements also. Therefore, the system class is defined by the highest order information-processing dynamic that it contains.     

This hierarchical view establishes the groundwork for describing successive classes of information processing, with learning and other agentic capacities representing potential extensions beyond the adaptive regime. In this framework, adaptivity refers to the spontaneous adjustment of a system’s internal dynamics in response to changing inputs or environmental conditions. So defined, adaptivity can be regarded as a minimal expression of agency, but it remains distinct from learning as defined in neuroscience and machine learning, where behaviour or parameters are modified through goal-directed optimisation or reward-based processes~\cite{friston_what_2012, kagan_harnessing_2025, miconi_neural_2025}. This perspective focuses on the intrinsic relationships among input, internal state, and output, rather than on external performance. Importantly, understanding the information-processing order of a system offers a principled way to identify the necessary system dynamic precursors to qualify for traits such as agency. Below, we now formally define these different orders of information processing. 

\begin{figure}[!h]
    \centering
    \includegraphics[width=0.5\linewidth]{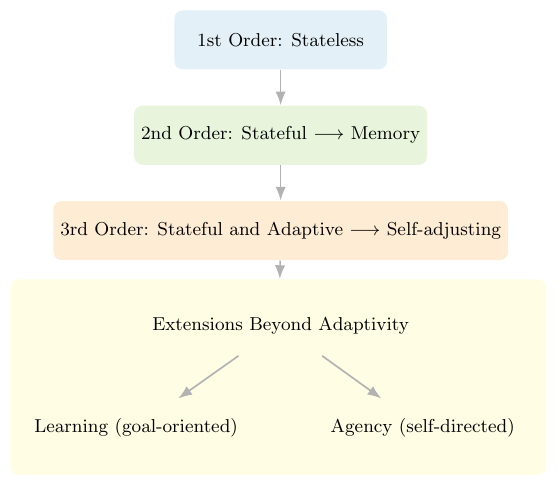}
    \caption{Information-processing classes and their relation to memory, adaptivity, and agency.}
    \label{fig:infoprocorder}
\end{figure}

\subsection{Class I: First-order information-processing}
Consider a system where information is received, progresses through the system, then triggers an action without any meaningful qualitative alterations to information quality so that, aside from some arbitrary noise term, given a knowledge of the system function the input would be derivable from the output minus the introduced noise. For example, a Rube Goldberg machine - no matter the complexity - does not qualitatively transform information. A process is initiated and continues through a set process according to non-dynamic rules such that the reverse of these rules would render any input invertible from the output. Such a system would undergo what we define as 'first-order information-processing' where it operates in a memoryless, reactive manner. Formally, we can represent the dynamics of a first-order information-processing system as:

\begin{align}
R(t)=\alpha(t)\,I(t)+\varepsilon(t), %\tag{1}\label{eq:trigger}
\end{align}
where $R(t)$ is the output of a system at time $t$, $I(t)$ is input from incoming information received at time $t$, $\alpha(t)$ is a time-varying gain or bias that is independent of $I(t)$ and $\varepsilon(t)$ is the zero-mean random error term. The mapping \(I(t)\mapsto R(t)\) is the identity up to a gain/bias \(\alpha(t)\) that may drift with time or other exogenous variables; it never depends on the specific value of \(I(t)\).  
Hence, the kind of information is preserved (no qualitative change), while the system may scale or offset it and add noise. This matches the idea that while information may have an effect on the system, the system's effect on information is fixed even if the trigger itself varies. Under this framework, a thermostat - even one that dynamically adjusts a threshold - has only this first-order information-processing. The thermostat receives information and, if a threshold is reached, acts upon that information. Even if several thresholds are required for the action to occur or several types of information are considered in this process, the information itself is not qualitatively changed by the system and the change itself is of a constant type. Biological systems may also operate with this first-order information-processing as this is the process observed in simple reflexes. 

%PCA or decision trees as other examples?

Examples of such systems can be decision trees, ohmnic resistors, linear time-varying systems with exogenous, time-varying modulation(LTV)~\cite{bittanti_periodic_2009}, and classical on/off thermostats~\cite{pickering_what_2024}.

%Give some examples of how this may be represented in more concrete terms such as look up tables, linear functions etc. 

% {\color{blue} Examples of such systems are: a resistor with linear, static gain; a linear time-varying system (LTV) with exogenous periodic gain modulation (either a square or a sinusoidal wave)~\cite{}; a classical on/off thermostat with time-varying gain coefficient~\cite{Periodic systems Bittanti-Colaneri, Nonreciprocal Gain in Non-Hermitian Time-Floquet Systems}.} % More information about these systems, and their information-processing order can be found in~\ref{sec:examples}}

\subsection{Class II: Second-order information-processing}
Consider another system where information is received by a system, where the system then acts upon the information to effect a qualitative, non-invertible change. This change may lead to alterations to the quality of the information, but the type of change itself is not amenable to change based on the information quality. This system can be considered stateful (holding a given state at a particular time), but it is not in-of-itself dynamically adaptive through its own internal mechanisms. Type I blindsight patients may be an example of this. These patients are able to respond to visual information, yet due to damage to the visual cortex, do so without any consciousness awareness. Previously, this phenomenon was discussed as an example of intelligent actions without consciousness \cite{kagan_neurons_2022}, but it may also represent a form of second-order information processing. In this case, visual information is received by the eye before being transmitted directly to the superior colliculus and lateral geniculate nucleus of the thalamus \cite{burra_affective_2019}. These neural systems act upon the received information, reorganizing and redistributing it through non-linear methods to trigger motor cortex activity that then drives a limited variety behaviour. Despite the qualitative transformation of information by lower brain regions, the qualitative change is itself deterministic based on preset criteria. This can drive a range of moderately complex responses and so are not reflexes, but may also not be considered agentic actions and indeed, those patients who respond based on Type I blindsight processes report having no awareness of self-perceived agency for the action \cite{brogaard_are_2011}. Formally, this process can be represented as: 

\begin{align}
R(t)=\mathcal{T}\!\bigl[I(t)\bigr]+\varepsilon(t) \label{eq:static}
\end{align}

which extends upon a first-order representation by the addition of a fixed (time-invariant) transformation operator $\mathcal{T}$. The operator \(\mathcal{T}\) can reshape, filter, threshold, or otherwise change the input qualitatively, but \(\mathcal{T}\) itself is constant—independent of the particular instance of \(I(t)\).  

Examples of this kind of systems are, RC low-pass filters, feedback control systems (such as a Bang Bang thermostats), and non-linear two-point circuit elements with memory, such as memristors~\cite{chua_memristor-missing_1971,caravelli_memristors_2018,caravelli_mise_2018}.

Therefore, the nature of the transformation never evolves based on the information incoming or the output, such that the change that occurs to the information does not itself change.

\subsection{Class III: Third-order (and higher) information-processing}

Consider a final exemplar system where information is received by the system, whereby the system then acts upon this information to effect a qualitative, non-invertible change. Yet, this change can itself be altered by the system's internal dynamics based on the previous outcome of what occurred in response to the system's previous behaviour following the previously received information. This allows the system to adaptive and self-modulating over time. Class III systems can display significantly more complex dynamics than Class I and Class II systems, as adaptivity dynamics can themselves be layered (increasing the order of information processing); however, by progressing to self-modulating, the system enters a distinct class lower order systems that are either memoryless or only stateful. Formally, the simplest class III system with a third-order information-processing relationship can be represented as:

\begin{align}
R_t &= \mathcal{T}_t\!\bigl[I_t\bigr]+\varepsilon_t,
\label{eq:adapt_a}\\[4pt]
\mathcal{T}_{t+1} &= \mathcal{G}\bigl(\mathcal{T}_t,R_t\bigr).
\label{eq:adapt_b}
\end{align}

which extends upon the representation of second-order information processing so that the effect of $\mathcal{T}$ on $I$ at time $t$ may differ over time according to the adaption rule $\mathcal{G}$ and to what the systems response ($R_t$) was at that time. The key additional here equation \eqref{eq:adapt_b} adds recursion: the transformation itself is a state variable updated after each output via \(\mathcal{G}\).  
Consequently, future inputs encounter a \(\mathcal{T}_{t+1}\) that depends on past outputs, capturing that the change itself may be changed by the output of the system for future and capturing adaptive behaviour.

Examples of such systems would match the existing modelled dynamics of biological neural networks \cite{viriyopase_when_2012}, along with artificial recurrent neural networks (RNN)\cite{tampuu_efficient_2018}, memristors with input-modulating mechanisms~\cite{caravelli_memristors_2018}, and Hebbian learning with Oja's rule (for stability )\cite{halvagal_combination_2023}. 

%Machine learning methods with backpropagation? Each node applies a transformation but based on backpropagation that transformation may vary over time.   

\maketitle

%%--------------------------------------------%%
\section{Case examples}
\label{sec:examples}
%%--------------------------------------------%%

Having outlined the theoretical basis of information processing classes, we now present some case examples corresponding to each class. These examples are not intended as exhaustive representations, but as minimal models that capture the essential features of each class: reactivity, state-based memory, and adaptivity. 

The three examples we will consider are: simple thermostat with a time-dependent exogenous gain (Class I with first-order); an ideal memristor~\cite{caravelli_memristors_2018,caravelli_mise_2018} (Class II with second-order); and an input-adaptive ideal memristor~\cite{caravelli_memristors_2018,caravelli_mise_2018} (Class III with third-order)
%%----
% \subsection{Examples of first complexity systems}

\subsection{First-order information-processing example: Thermostatic switch} 
In this example, the thermostatic switch modulates the system's gain over time according to a binary switching signal $s(t)$. The instantaneous gain $\alpha(t)$ is defined as 
\begin{align}
    \alpha(t)
=
    \alpha_\mathrm{off}
    \left[
        1
    -
        s(t)
    \right]
    + 
    \alpha_\mathrm{on}
    s(t),
\end{align}
so that the system alternates between two constant gain values, $\alpha_\mathrm{off}$ and $\alpha_\mathrm{on}$, corresponding  to Off and On states of the switch. Although the overall gain $\alpha(t)$ varies in time, the switching process is entirely exogenous, that is $s(t)$ evolves independently of the input signal $I(t)$. In this implementation, $s(t)$ is defined as a binary square-wave
\begin{align}
    s(t)
=
    H
    \left[
        \sin (2 \pi f_s t + \phi)
    \right],
\end{align}
where $H[\cdot]$ is the Heaviside step function. The sine modulation with frequency $f_s$ and phase $\phi$ produces a periodic square wave of amplitude $\{0,1\}$, and $50 \%$ ``duty cycle'', driving the system to alternate regularly between its On and Off states.

\begin{figure}[h!]
    \centering
    \includegraphics[width=0.8\linewidth]{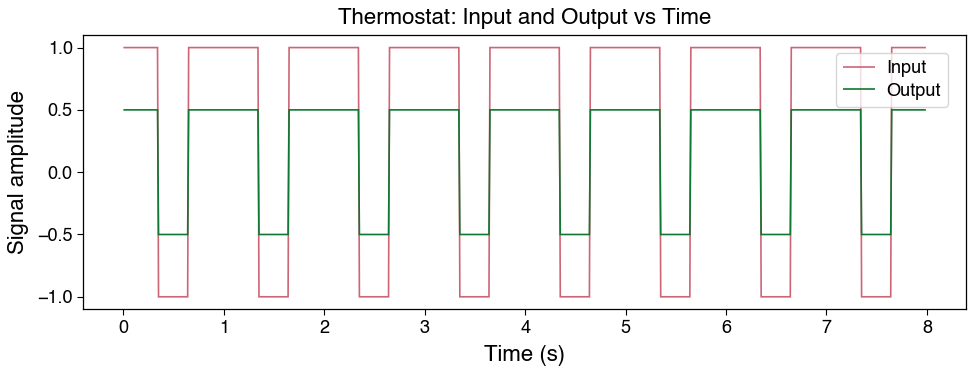}
    \caption{First-order thermostat response to a square wave: instantaneous, proportional switching between two fixed output levels; no memory or adaptation.}
    \label{fig:thermostat_I_O_time}
\end{figure}

To make the characteristics of this class of information processing more apparent, in Fig.~\ref{fig:thermostat_I_O_time} we show the input and output signals of the thermostatic switch as functions of time.
The input consists of a square wave that alternates periodically between positive and negative amplitudes, driving the system through successive On and Off states.
Because the modulation of the gain $\alpha(t)$ is entirely exogenous, the output follows the input instantaneously, switching proportionally between two constant levels corresponding to $\alpha_\mathrm{on}$ and $\alpha_\mathrm{off}$.
Each change in input polarity produces an immediate inversion of the output, with no residual transients or delay once the input returns to zero.
% This behaviour illustrates the defining property of first-complexity systems: an instantaneous, memoryless transformation in which the input–output relation remains invertible at every point in time.
%

% To illustrate that this system is first complexity, we consider a linear input
% %
% \begin{align}
%     u(t)
% =
%     a \, t
% \end{align}

%We have considered three first-complexity information processing systems, whose response  $R(t)$ is given in~\eqref{eq:trigger}, with input-independent gain: 
%
% \begin{enumerate}
% %
% \item A simple resistor, with static gain parameter $\alpha$;
% \item A linear time-varying system with exogenous, periodic modulation of the form
% %
% \begin{align}
%     \alpha(t)
% = 
%     \alpha_0 
%     \left[1 + a \sin (2 \pi f t) \right]
% \end{align}
% %
% %
% \item A thermostat switch, with time-dependent exogenous modulation
% %
% \begin{align}
%     \alpha(t)
% =
%     \alpha_\mathrm{off}
%     \left[
%         1
%     -
%         s(t)
%     \right]
%     + 
%     \alpha_\mathrm{on}
%     s(t).
% \end{align}
% % 
% The switching signal $s(t)$ is a binary square-wave
% %
% \begin{align}
%     s(t)
% =
%     H
%     \left(
%         \sin (2 \pi f_s t + \phi)
%     \right)
% \end{align}
% %
% with frequency $f_s$ and amplitude $\{0,1\}$ (at $50 \%$ duty cycle). 
% \end{enumerate}

%%----
\subsection{Second-order information-processing example: Ideal memristor}
%%----

As an example of a second-order information processing system, we consider a \textit{memristor}, a two-terminal electrical component whose instantaneous resistance $\mathcal{R}(t)$ depends on the voltage or current previously applied to it~\cite{chua_memristor-missing_1971}: $V(t) = \mathcal{R}(\eta(t))  i(t)$, $V(t)$ is the voltage, and $i(t)$ is the current. Unlike a linear resistor, whose resistance is fixed, a memristor retains a memory of past inputs through an internal state variable $\eta(t)$ that modulates its resistance. The device thus combines instantaneous response and temporal integration within a single element, naturally implementing a system with internal memory. In particular, we will consider an ideal memristor~\cite{chua_memristor-missing_1971, caravelli_mise_2018}, whose resistance is given by
\begin{align}
    \mathcal{R}(\eta(t))
=   
    \mathcal{R}_\mathrm{off}
    \left(
        1
    -
        \eta(t)
    \right)
    +
    \mathcal{R}_\mathrm{on}
    \eta(t), 
\end{align}
where $\mathcal{R}_\mathrm{off}$ and $\mathcal{R}_\mathrm{on}$ denote the high and low-resistance limits of the device, respectively.  The system's dynamics follow the equations of motion of the internal state variable
\begin{align}
    \frac{d\eta}{dt}
=
    \frac{\mathcal{R}_\mathrm{off}}{\beta}
    \frac{V(t)}{\mathcal{R}(\eta(t))}
-
    \alpha \, 
    \eta(t).
    \label{eq:memristor}
\end{align}
where $\alpha$ determines the relaxation rate of the internal state and $\beta$ is an effective activation voltage sets the coupling between the applied voltage $V(t)$ and the memristor's internal dynamics.
The first term drives the state in response to the applied voltage, while the second ensures exponential relaxation toward equilibrium in the absence of input.
As a result, the current response of the system depends both on the instantaneous voltage and on the memory encoded in $\eta(t)$, which integrates the effect of past inputs.
It is important to note is that for a voltage-driven ideal memristor, the incoming information $I(t)$ can be encoded in the voltage $V(t)$, while the response $R(t)$ corresponds to the output current $i(t)$. Therefore, in the same notation of eq.~\eqref{eq:static}, the fixed transformation operator is defined as
\begin{align}
    \mathcal{T}[V(t)]
=
    \frac{V(t)}{\mathcal{R}(\eta(t))}
\end{align}
Although it may seem that the operator $\mathcal{T}[V(t)]$ is time-dependent via $\eta(t)$, its functional form and governing parameters remain fixed. Once the parameters $\alpha$, $\beta$, $R_\mathrm{off}$ and $R_\mathrm{on}$ are fixed, the memristor’s transformation law remains fixed in form. In this sense, the ideal memristor implements a fixed operator with state-dependent dynamics (not input dependent). 
This example illustrates how second-order information-processing systems differ qualitatively from first-order ones: they transform the input through a fixed operator whose instantaneous response depends on an internal variable carrying information about the past. 

\begin{figure}[h!]
    \centering
    \includegraphics[width=0.8\linewidth]{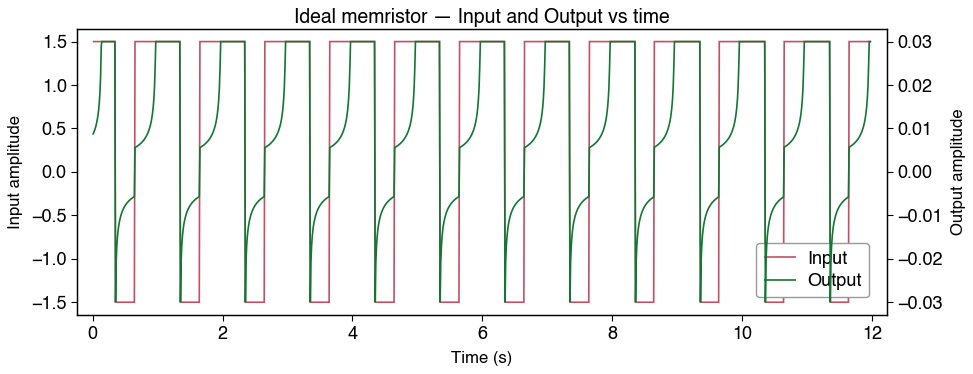}
    \caption{Ideal memristor response to a square wave. The output current reflects the slow evolution of the internal state, so each input transition leaves a carry-over in the next cycle; the mapping is fixed, but the response depends on recent history.}
    \label{fig:memristorNA_I_O_time}
\end{figure}

Similarly to the thermostatic switch, we consider the response of this system to a bipolar square-wave input, shown in Fig.\ref{fig:memristorNA_I_O_time}. In this case, while the input switches sharply between positive and negative amplitudes, the output current responds more gradually, reflecting the slow evolution of the internal state variable $\eta(t)$. Each voltage transition leaves a transient trace in the output, producing curved segments that persist beyond the moment of input reversal. This delayed and history-dependent response demonstrates that the memristor integrates information over time: although its transformation law remains fixed, the instantaneous output depends not only on the present input but also on the residual state established by previous inputs.

% \begin{enumerate}
%     \item Fixed low-pass RC filter 
%     \item Thermostat with bang-bang switch (or Schmitt trigger), that is now a function of the input. 
%     \item Memristive device (memory but no adaptability)
% \end{enumerate}

\subsection{Third-order information-processing example: Memristive Bioreceptor}

%To extend the ideal memristor into a third-complexity information-processing system, 
Following bio-receptor adaptation, we define a third-order information-processing system  as an ideal memristor with a slow state-dependent mechanism that tunes the effective input gain and bias in response to recent activity. We refer to this device as a memristive bioreceptor (MBR), by which we mean a memristive element that implements receptor-style adaptation of input gain and bias via a single internal state (fading memory) and slow observer variables; no molecular binding or learning is implied. The memristor’s internal state provides the system’s intrinsic (fading) memory, while the slow mechanism modulates sensitivity without learning. We add slow, adaptive mechanisms, akin to neuronal homeostasis, that adjust the input amplitude and bias from slow averages of past activity~\cite{ladenbauer_how_2014, benda_neural_2021}. In the present system, adaptation is implemented through exponentially weighted moving averages (EWMAs) that provide smoothed estimates of internal variables over time-scales much longer than the intrinsic device dynamics~\cite{ladenbauer_impact_2012}. These averaged quantities act as ``observer'' variables that modulate the device’s transformation operator, enabling it to adjust its own amplitude and bias in response to sustained changes in internal or external activity.
At each time step, the applied voltage is constructed as
\begin{align}
    V(t) = A(t)\,u(t) + b(t),
    \label{eq:mem_adaptive_input}
\end{align}
where $u(t)$ is the external input signal, while the adaptive variables $A(t)$ and $b(t)$ respectively modulate the effective input amplitude and bias. Their dynamics evolve on a slower timescale than the internal state of the memristor, according to the update rules
\begin{align}
    \frac{dA(t)}{dt} &= \gamma_A \big[\sigma_\phi^2(t) - \sigma_*^2\big], \\
    \frac{db(t)}{dt} &= \gamma_b \big[\eta_* - \bar{\eta}(t)\big],
\end{align}
Here, $\bar{\eta}(t)$ and $\sigma_\phi^2(t)$ denote EWMAs of the internal state $\eta(t)$, and of its activity measure~$\phi(t) = \eta(t)\left(1-\eta(t)\right)$, and are defined, respectively as
\begin{align}
    \frac{d\bar{\eta}(t)}{dt}
=
    \frac{1}{\tau_\eta}
    \left[
        \eta(t)
    -
        \bar{\eta}(t)
    \right],
    \qquad
    \frac{d\sigma^2_\phi}{dt}
=
    \frac{1}{\tau_\phi}
    \left[
        \left(
            \phi(t)
        -
            \bar{\phi}(t)
        \right)^2
        -
        \sigma^2_\phi
    \right].
\end{align}
The parameters  $\tau_\eta$ and $\tau_\phi$ are adaptation time constants, setting the rate at which $\bar{\eta}$ and $\sigma^2_\phi$ integrate past activity. The parameters $\gamma_A$ and $\gamma_b$ control the adaptation rate of amplitude and bias, while $\eta_*$ and $\sigma_*^2$ set target values for the long-term mean and variability of the internal state. The amplitude adaptation thus responds to sustained deviations in the variability of internal activity via $\sigma^2_\phi$, whereas the bias adapts to deviations in its mean level $\bar{\eta}(t)$, allowing the system to maintain stable yet responsive behaviour under changing input conditions. 
The implementation of EWMAs is similar in spirit to biophysical models where adaptation currents act as low-pass filters of recent activity, effectively implementing exponential averaging over time~\cite{ladenbauer_how_2014, benda_universal_2003, benda_neural_2021}.

\noindent In the notation of Eqs.~\eqref{eq:adapt_a} and~\eqref{eq:adapt_b}, the input transformation rule and the adaptation map are
\begin{align}
    \mathcal{T}_k [V_k]
=
    \frac{A_k u_k + b_k}{\mathcal{R}(\eta_k)},
    \qquad
    \mathcal{G}
    \left[
    \mathcal{T}_k, R_k
    \right]
=
    \left\{
    \begin{aligned}
        \dot{A}_k \\
        \dot{b}_k
    \end{aligned}
    \right.
\end{align}

Through these coupled adaptive feedbacks, the memristor no longer operates with a fixed transformation law,  but continuously reconfigures its input–output mapping in response to its own activity. The internal state $\eta(t)$ provides a short-term trace of recent inputs (short term memory), while the adaptive variables introduce a slower, longer-term memory that regulates the device parameters. In this sense, the device exhibits self-modulating dynamics, where both memory and adaptive regulation jointly determine the transformation of incoming signals.

\begin{figure}[h!]
    \centering
     \includegraphics[width=0.8\linewidth]{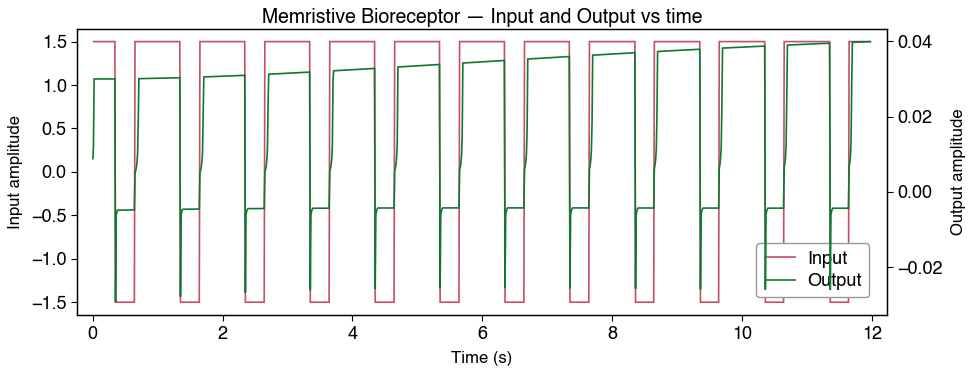}
    \caption{Memristive bioreceptor under square-wave drive. The output exhibits memristive transients and gradual changes in sensitivity and offset driven by slow activity averages, indicating that the input–output mapping is itself adapting over time.}
    \label{fig:memristorA_I_O_time}
\end{figure}

Similarly to the previous two examples, we consider the response of this system to a bipolar square-wave input, shown in Fig.~\ref{fig:memristorA_I_O_time}. As for the ideal memristor, the output current activates with a slight delay. However, it also gradually shifts in both amplitude and baseline across successive cycles. Each transition retains the characteristic curvature of the ideal memristor, albeit not as pronounced; however, the overall waveform drifts upward as the system slowly adjusts its effective gain and bias through the adaptive variables$A(t)$ and~$b(t)$. These slow, cumulative adjustments show that the transformation governing the input–output relationship is no longer fixed, but evolves in response to the system’s own ongoing activity.
The device therefore combines short-term memory with a slower, adaptive modulation of its transformation law.

%%----------------------------------------------
\section{Characterisation of memory and adaptivity}

%%----------------------------------------------

To further characterise the emergence of memory and adaptivity across the three systems, we examine their dynamics in the input-output plane (IO plane), which capture how each system retains information about past inputs, and through simple temporal measures of the input–output relationship, such as sliding-window linear regression and zero-crossing lag analysis, which reveal gradual changes in effective gain and bias indicative of adaptive dynamics.

%%----------
\subsection{Input-Output trajectories and memory}
%%----------
To illustrate how memory emerges as the information-processing order increases, we analyse each system's trajectories in the input-output (IO) plane under periodic driving, and determine whether it forms hysteretic trajectories. The presence of hysteresis loops is taken as a signature of memory, following the interpretation established in memristive devices, where hysteresis in the current-voltage (I-V) characteristic is considered a hallmark of memristive behaviour, and more generally of memory~\cite{chua_memristor-missing_1971,pershin_memory_2011}

More broadly, the association between hysteresis and memory has deep roots in condensed matter physics and extends across many driven and disordered systems, from magnetic and ferroelectric materials to amorphous solids and other non-equilibrium media, where hysteresis responses give rise to return-point memory, or history-dependent effects~\cite{sethna_hysteresis_1993,keim_memory_2019}. 

\begin{figure}[!h]
    \centering
    \includegraphics[width=1.0\linewidth]{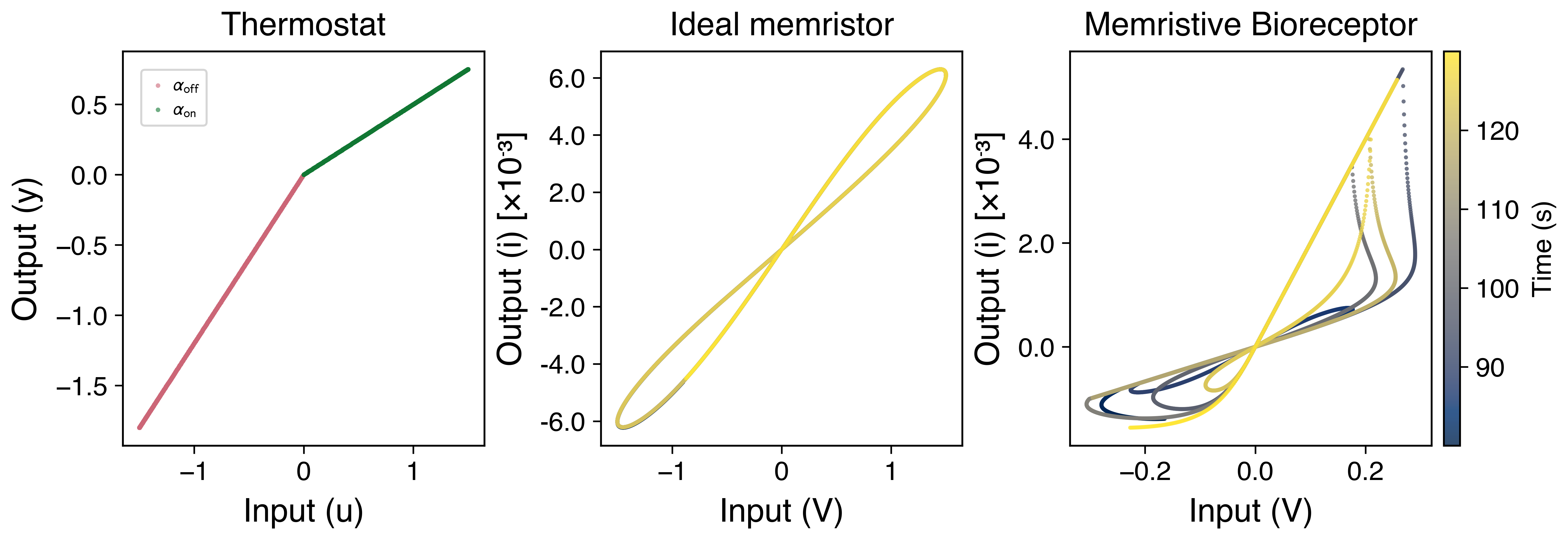}
    \caption{Input-Output trajectories for the three example systems, all driven by a sinusoidal input. The thermostat (left) produces two fixed gain lines corresponding to the constant gain values $\alpha_\mathrm{on}$ and $\alpha_\mathrm{off}$ (On and Off state), showing an instantaneous, memoryless response. The ideal memristor (centre), exhibits a stationary voltage-current loop, where points at earlier times (darker blue) are retraced by later ones (lighter yellow), indicating memory without adaptivity. The MBR (right) shows a loop that gradually shifts and deforms over time, (colour scale indicates absolute time within the final 50 s of simulation), demonstrating memory with adaptivity}
    \label{fig:phasetrajectory_compare_all_models}
\end{figure}

In Figure~\ref{fig:phasetrajectory_compare_all_models}, we show the trajectories of the three examples systems, considered in section~\ref{sec:examples}, in the IO plane when driven by a sinusoidal input. In the thermostat (left), the input represents a dimensionless driving signal, while the output corresponds to the state of the instantaneous state of the system. The resulting points cluster along two points associated with the two constant gain values $\alpha_\mathrm{on}$ and $\alpha_\mathrm{off}$ (corresponding to the On and Off state of the switch), indicating an instantaneous, memoryless mapping between input and output.

For the memristive systems (centre and right) in \ref{fig:phasetrajectory_compare_all_models}, the input corresponds to the applied voltage $V(t)$ and the output to the resulting current $i(t)$, for the final 50 seconds of simulation ($t = 80\text{–}130~\mathrm{s}$). After transient behaviour has decayed, the colourmap indicates absolute time within this window, showing how the adaptive device’s response evolves. In the ideal (non-adaptive) memristor (centre), the voltage-current trajectory forms a stationary loop, as indicated by the fact that points from earlier times (shown in darker blue) are retraced and covered by more recent ones. This demonstrates, as expected for an ideal memristor, that the current depends on both the present voltage and the internal state established by previous inputs, reflecting a fixed transformation that nonetheless retains memory of past activity, that is memory without adaptivity.

In the memristor bio-receptor (right), the current-voltage trajectory no longer retraces a stationary path. Instead, successive segments of the loop shift progressively over time, as shown by the change in colour from darker to lighter shades along the curve. This drift indicates that the relationship between the voltage and the current evolves as the system's adaptive variables ($A(t)$ and $b(t)$, introduced in Eq.~\eqref{eq:mem_adaptive_input}) adjust to current and previous activity. The gradual deformation and displacement of the loop therefore qualitatively signifies memory with adaptivity, that is a transformation that not only depends on past inputs but also self-modifies through slow parameter evolution.

% \begin{figure}
%     \centering
%     \includegraphics[width=0.7\linewidth]{Figures/memristors_windows_last.png}
%     \caption{Enter Caption}
%     \label{fig:memristors_window}
% \end{figure}

% In Figure~\ref{fig:memristors_window}, the difference between the dynamics of the two systems  

It is important to add that, while the presence of hysteresis in the IO plane provides a qualitative indication of memory, it does not directly quantify its magnitude or timescale. 
Different loop shapes may reflect distinct underlying dynamics, such as saturation, non-linearity, and adaptation, without allowing straightforward measures of how much past information is retained. 
Quantitative assessment of memory typically requires complementary temporal analyses based on temporal correlations or information-theoretic measures, such as output-signal autocorrelation or information storage~\cite{lizier_local_2012}.

%%----------
\subsection{Rolling regression, zero-crossing lag and adaptivity}
\label{sec:rollfit}
%%----------

Although we do not seek to quantify adaptivity precisely, we can still infer meaningful information from two simple temporal analyses: rolling (sliding-window) regression~\cite{zivot_rolling_2006} and zero-crossing lag analysis.

\begin{figure}[!h]
    \centering
    \includegraphics[width=0.75 \linewidth]{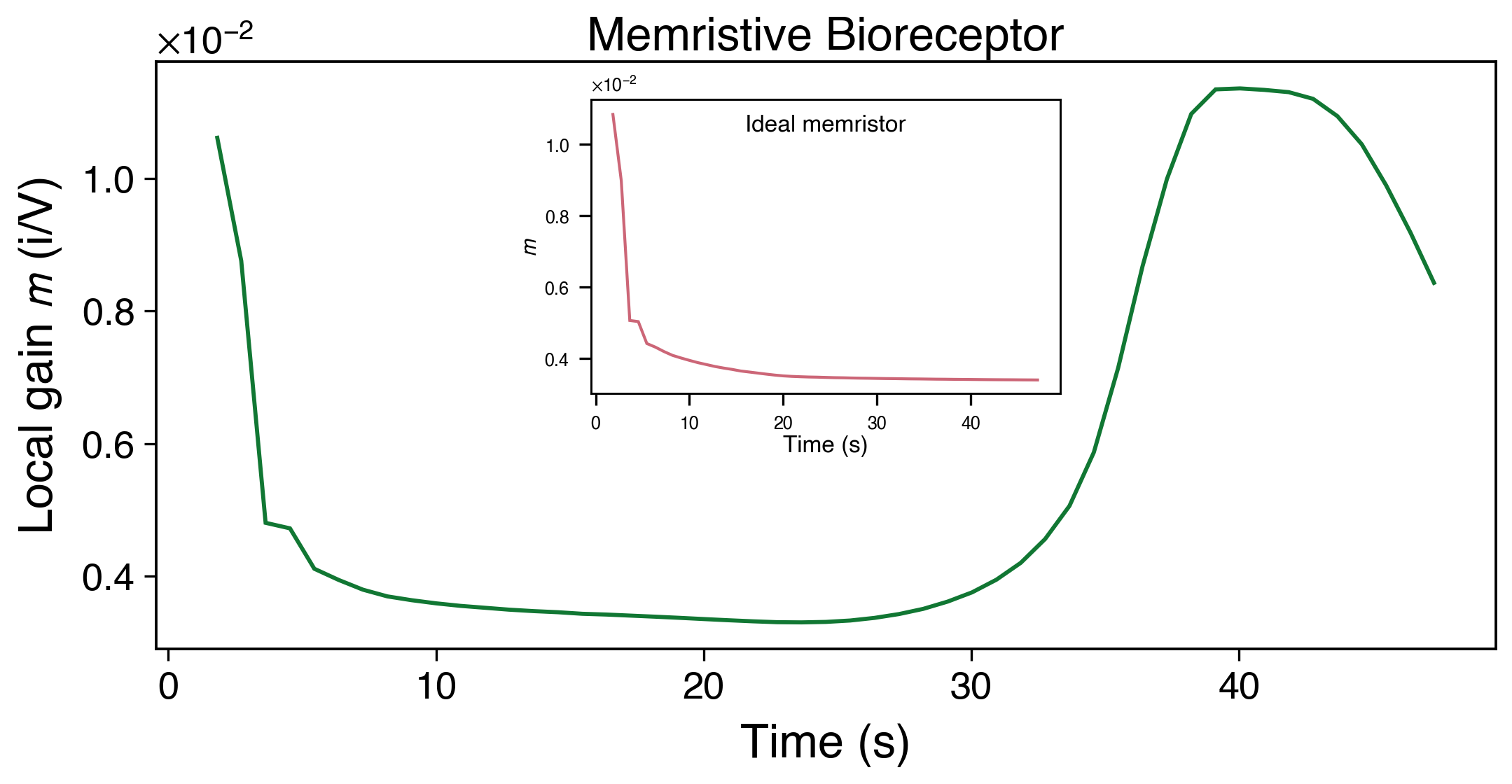}
    \caption{Local gain $m(t)$ obtained from sliding-window regression under a slow-varying sinusoidal input. The memristive bioreceptor (main panel) shows a clear frequency-dependent increase in gain. Inset: corresponding $m(t)$ for the ideal memristor, which remains approximately constant.}
    \label{fig:rolling-fit-adaptive-non-adaptive-gain}
\end{figure}

Rolling regression captures slow variations in the strength and baseline of the response by fitting a linear model to short, overlapping segments (or sliding windows) of the input–output data. This provides  local (in time) estimates of the system's effective gain and bias, quantities that describe how strongly the system reacts to its input and around which baseline level. As shown in Figure \ref{fig:rolling-fit-adaptive-non-adaptive-gain}, by observing how the local gain factor evolves over time, we can assess whether the system’s transformation operator remains fixed or changes dynamically: if the local gain is constant or independent of the input, the system behaves as a first-order processor, whereas systematic variations would reveal the presence of memory or adaptation\footnote{Conceptually, this approach is related to local or weighted least-squares estimation methods used to track time-varying parameters in adaptive systems~\cite{joensen_tracking_2000}, but here it is implemented explicitly through a finite moving window rather than recursive weighting.}.

The zero-crossing lag focuses instead on response timing,  comparing when the input and output cross a reference level (typically zero, but more generally, any baseline such as the mean or steady-state value). A constant lag suggests stationary response, while shifts in lag over time can be interpreted as adaptive changes in temporal alignment. 

Together, these two measures capture complementary aspects of systems' behaviour: one describing how the amplitude and offset of the response evolve, while the other shows how the timing of the system's response adjusts relative to the driving signal.

We have applied sliding regression to the three systems we have considered so far: thermostat, ideal memristor and memristive bioreceptor. As a probing signal, we have used a slow-varying sinusoid input, or ``chirp'', whose instantaneous frequency sweeps smoothly from $0.1\;\si{Hz}$ to $10\;\si{Hz}$. 

After selecting a window of appropriate length (see below), rolling regression is obtained by considering a local (window-sized) linear input-output approximation given by
\begin{align}
    y(t) 
\approx 
    m(t)\,u(t) 
+   
    b(t),
\end{align}
where $m(t)$ captures the instantaneous sensitivity (or gain) of the system to its input, and $b(t)$ represents the locally inferred baseline around which the output fluctuates.

Frequency-swept driving has been used to reveal how non-linear dynamical systems change their response as the forcing frequency varies~\cite{hudson_measuring_2012}. Here it serves the same role: showing how each system’s effective gain and bias evolve across different input timescales.

The sliding-window size was chosen relative to the slowest component of the input. Because the slow-varying sine wave begins at $f_0 = 0.1\ \si{Hz}$ (period $T_0 \approx 10\ \si{s}$), we used a window $\mathrm{W} = 0.4 \ T_0 \approx 4\ \si{s}$ with a step $\mathrm{S} = 0.1 \ T_0 \approx 1\ \si{s}$. This window is short enough to resolve slow drifts in the system’s effective gain and bias, yet long enough for each local linear fit to be numerically stable. The same window parameters were used for all systems to ensure comparability. The step size was set to one quarter of the window length, which provides enough overlap (approximately $75\%$) for stable estimates, while allowing the regression window to move gradually across the signal.

 \begin{figure}[h!]
    \centering
    \includegraphics[width=0.8\linewidth]{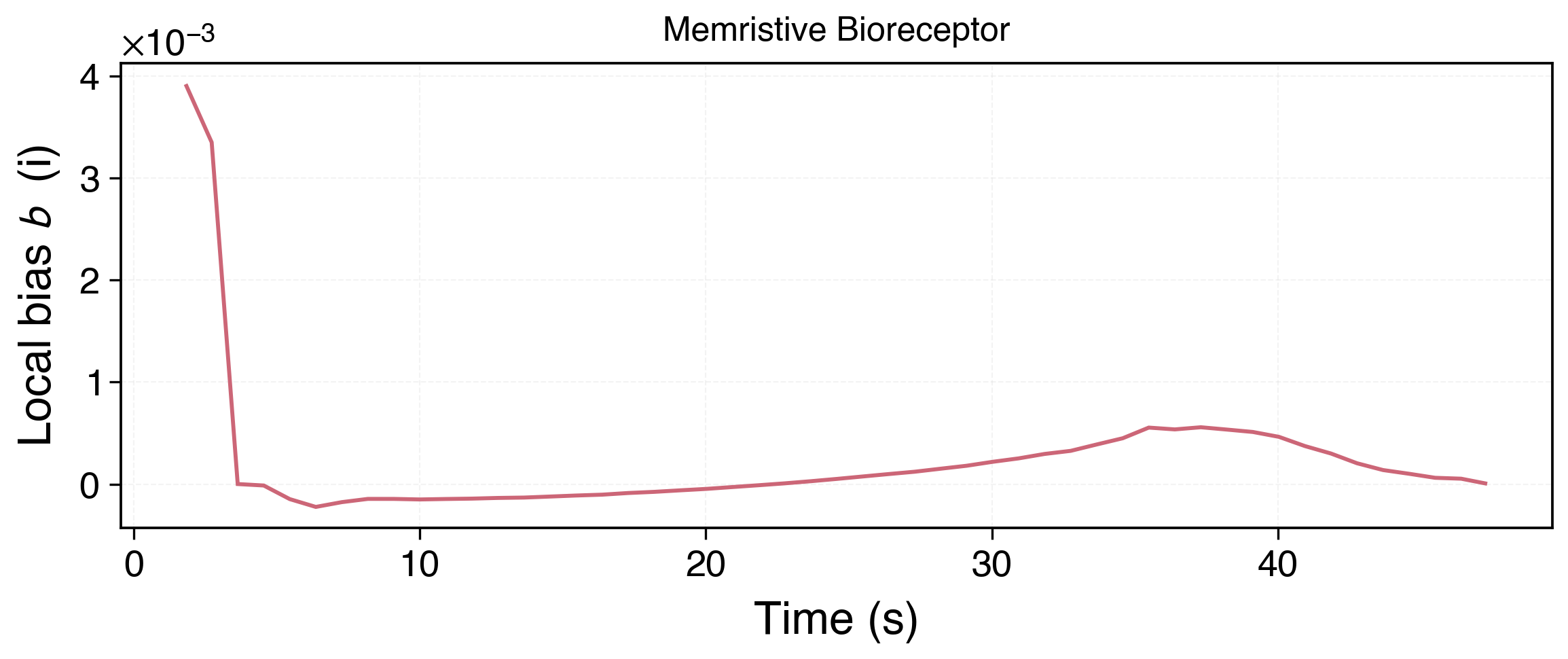}
    \caption{Local bias $b(t)$ for the MBR obtained from sliding-window regression. The bias exhibits slow drift consistent with the device’s time-dependent internal state.}
    \label{fig:rolling-fit-adaptive-bias}
\end{figure}

Figure~\ref{fig:rolling-fit-adaptive} in Appendix~\ref{sec:app_roll_regression} shows the evolution of the local gain $m(t)$ for the memristor bioreceptor. The initial portion (up to $t \approx 5 \ \si{s}$) displays the expected decay associated with the memristor’s state relaxation, but as the input frequency increases, the device progressively amplifies the signal, leading to a clear rise in the gain around $t \approx 30 -40\,\si{s}$. For comparison, the inset figure shows the corresponding $m(t)$ for an ideal (non-adaptive) memristor under the same input: after an initial transient the local gain simply relaxes to a constant value and remains flat, illustrating that the adaptive behaviour observed in the main panel is not a generic property of memristors, but rather arises from the additional adaptive state variable.

Figure~\ref{fig:rolling-fit-adaptive-bias} shows the evolution of the local bias $b(t)$. The bias rapidly settles near zero and then drifts upward as the device adapts, before declining again once the state begins to saturate. This behaviour is again absent in the ideal memristor and in the thermostat system (see Figures~\ref{fig:rolling-fit-thermo},\ref{fig:rolling-fit-nonadapt}, in appendix~\ref{sec:app_roll_regression}), both of which maintain nearly constant bias throughout.
The memristive bioreceptor exhibits characteristic signatures of time-dependent internal dynamics: both the local gain and local bias change over the course of the experiment in a manner that reflects slow relaxation at early times followed by a pronounced gain increase as the input frequency becomes sufficiently high to activate its adaptation mechanism.

To complement the amplitude-based analysis, we briefly examine the response timing using the zero-crossing lag. For this purpose, we have used a bipolar square-wave input (similar to that introduced in Figures.~\ref{fig:thermostat_I_O_time},~\ref{fig:memristorNA_I_O_time},~\ref{fig:memristorA_I_O_time}), since its abrupt sign changes provide unambiguous temporal reference points. 

To quantify the timing, we extract the zero-crossing times of both the input and output (using a threshold of zero, which is natural in this case as the bipolar square wave has mean zero). Each input sign change yields a time $t^k_{u}$, and the corresponding output time change (obtained by linear interpolation) gives $t^k_{y}$. The lag at each $k$-crossing  is given by:
\begin{align}
    \ell^{(k)} 
= 
    t^{(k)}_y - t^{(k)}_u
    ,
\end{align}
with positive values indicating that the output trails the input, and negative values indicating a lead. 

\begin{figure}[!h]
    \centering
    \includegraphics[width=0.75\linewidth]{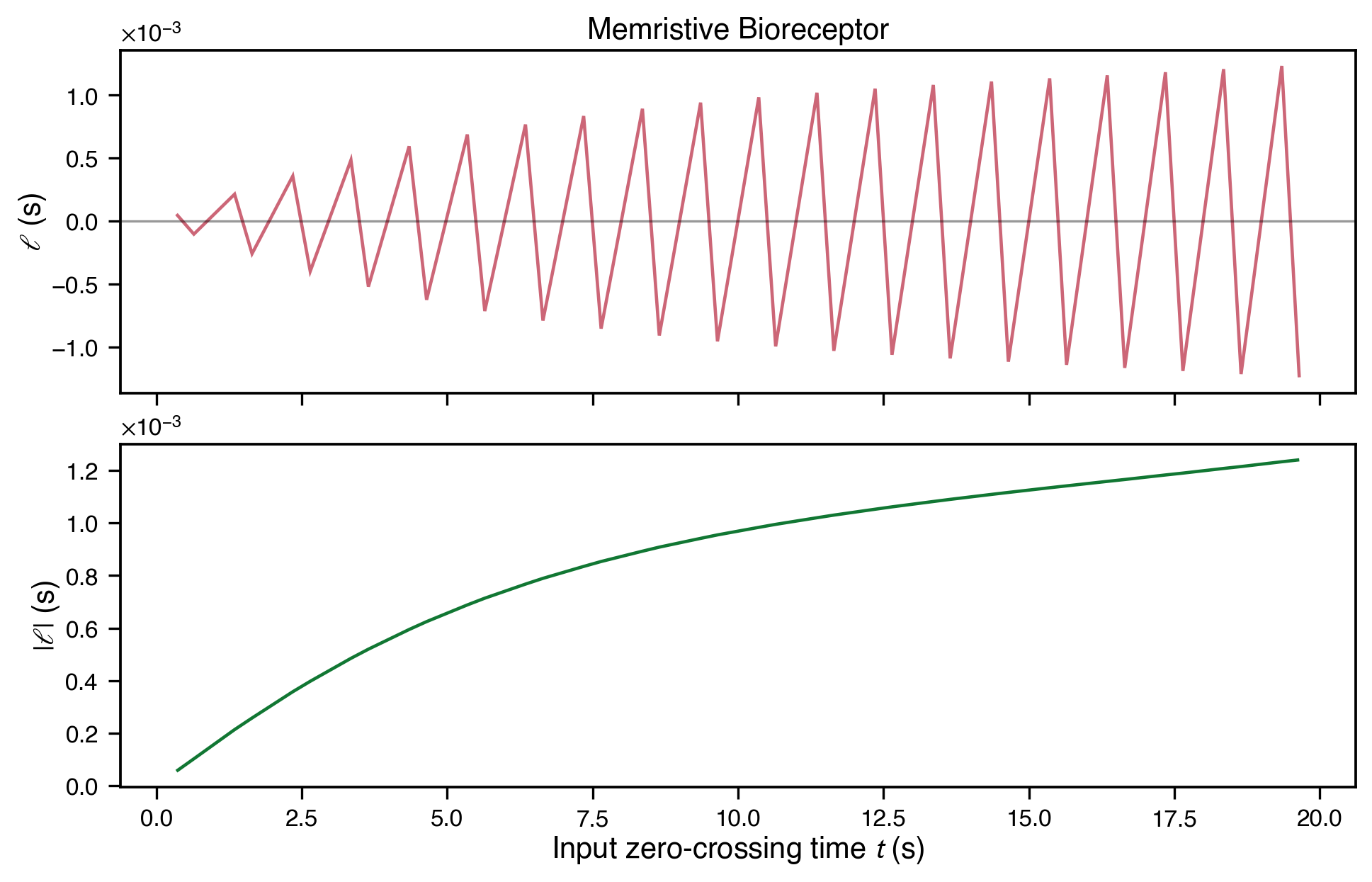}
    \caption{Zero-crossing lag  for a MBR evaluated at each input zero-crossing time $t^{(k)}_u$. Top: signed lag $\ell^{(k)}$ shows alternating polarity due to the bipolar drive, with a growing amplitude over time. Bottom: the absolute lag $|\ell^{(k)}|$ increases steadily, revealing a gradual timing drift driven by the system’s adaptive dynamics.}
    \label{fig:zerocrossinglag_adaptive}
\end{figure}

In Figure~\ref{fig:zerocrossinglag_adaptive}, the absolute lag $|\ell(t)|$ shows a numerically small but steadily increasing drift. The $x$-axis shows the input zero-crossing times $t^{(k)}_u$, since the lag is defined only at those instants; using $t^{(k)}_u$ places each measurement at the exact moment the input flips sign, ensuring the lag is referenced to the correct generating event. The lag $\ell(t)$ (top panel) oscillates in sign, as expected from a bipolar square-wave input, but the amplitude of these oscillations grows over time, indicating that the output begins to deviate from perfect synchronicity with the input. The bottom panel makes this trend explicit: $|\ell^{(k)}|$ increases monotonically across consecutive crossings. For reference, the corresponding lag curves for the thermostat and ideal memristor, presented in appendix~\ref{app:zerocrossing}, display a different behaviour. The thermostat retains zero lag (Figure~\ref{fig:zerocross-thermo}) and although the ideal memristor shows a small, fixed, time offset (Figure~\ref{fig:zerocross-ideal}), the magnitude of its oscillations remains constant over time. The slow growth of $|\ell|$ is a clear signature of time-dependent internal dynamics.

% Because the bipolar input is symmetric, with positive and negative phases of equal duration, any fixed delay introduced by the system affects all crossings equally in both directions. This makes the mean lag close to zero for all three systems—what matters is not the average delay, but whether the lag changes over time as the system evolves. A non-adaptive system should produce fluctuations around zero with no systematic trend, whereas an adaptive system may show a drift or systematic modulation in \ell^{(k)}.

% The x-axis of the lag plots shows the input zero-crossing time, t_u^{(k)}, rather than ordinary time.
% This simply labels each lag value by the exact moment of the corresponding input sign change, giving a clean and direct time axis for the discrete sequence \ell^{(k)}.
% In other words, each point represents “the lag at this particular transition of the input,” allowing us to visualise how timing alignment evolves across the input sequence.

%%----------
\section{Applicability of the Information-Processing Order Framework}
The utility of the presented framework is as a substrate-independent dynamic that can be tested in a range of systems. This does not aim to replace other considerations that may relate to information-processing architecture, but it does provide a measurable system level computational mechanism that can be assessed. This fits neatly with common frameworks that breakdown considerations into levels such as computational, algorithmic, and implementational/physical \cite{marr_vision_2010}. By isolating a core computational property, this is a particularly useful way to define whether a system displays a trait such as agency. We propose that there are three broad areas that this framework will benefit: ethical considerations, functional potential, and substrate independent comparisons. 

\subsection{Ethical Considerations of Agency}
The nature of agency plays an important role in ethics, especially in the context of free will, moral responsibility, and moral status \cite{brey_moral_2014}. Moral status is the concept that an entity matters in and of itself, that it has interests. Moral status is the basis of rights \cite{clarke_rethinking_2021}. Some philosophers, such as Gosepath (2014), consider agency to be a core component of moral status \cite{steinhoff_all_2015}. However, moral status is generally considered to require consciousness and so the kind of agency which would bestow moral status would be conscious agency. The challenge, however, has been the development of methods for identifying conscious agency, as a component of moral status, in non-paradigmatic cases (e.g. embodied brain organoids, artificial intelligence, neuroAI) where our proxies for inferring intentionality cannot be readily utilized. Thus, the development of reliable metrics for inferring whether an entity possesses morally relevant cognitive capacities, such as conscious agency, has become a core focus of experimental neuroethics \cite{kagan_embodied_2024}. Previous studies have devised strategies for inferring moral status-conferring cognitive capacities based on comparative behavioural responses \cite{boyd_dimensions_2024} or the neural architecture of information flow \cite{boyd_moral_2024}. These efforts attempted to ground moral consideration on epistemic criteria. The formalism proposed in this study provides a theoretical foundation for inferring cases of ‘minimal agency’, which we propose are antecedent to the kinds of mental representation typically associated with moral standing, that is, conscious agency  \cite{barandiaran_defining_2009}. The qualities associated with Class III agency (e.g. temporality, adaptability) are likely to be necessary but insufficient for manifesting moral status-conferring capacities, such as the capacity for self-directedness (or autonomy) and self-consciousness needed to have an enduring concept of oneself \cite{block_confusion_1995} or ability to form a conception of the ‘good life’ \cite{rawls_theory_2003}. Moreover, the aspects of agency that confer moral status, including consciousness, personal autonomy, and intentionality, are morally consequential in the extent to which they reflect an entity’s evaluative stance, the ability to have preferences, interests, or values that guide action \cite{blumenthal-barby_end_2024}. In brief, the possession of minimal agency, even Class III agency, does not necessarily confer moral status, but may be essential to understanding the causative mechanisms that give rise to morally-relevant manifestations of agency.  

\subsection{Functional Potential of Systems}
Agency is regularly positioned as a key feature of intelligent systems. Currently attempts to build intelligent systems are at an all-time high, most commonly with silicon based computing, but also with biological components \cite{betz_why_2025}. Being able to predict (or at least constrain) the range of behaviours that a system may exhibit that could ultimately create an impact based on the way it is able to process information would be useful. This impact-focused approach is also consistent with other calls to define the agency of a system based primarily on how the system can impact the information environment it is embedded within \cite{soltanzadeh_metaphysical_2025}.  One particular area where proposed information-processing order framework can be useful is in the rapidly growing field aiming to use living biological neural cultures for information processing and eventually intelligent purposes \cite{kagan_harnessing_2025}. While the number of tools designed to interact with neural cultures have been growing \cite{kagan_cl1_2025, jordan_open_2024, zhang_mind_2024, cai_brain_2023}, there is still considerable uncertainty over how to assess the tasks assigned to these neural cultures \cite{tanveer_starting_2025}. To further complicate matters, there are diverging efforts on how to best structure neural cultures for the purpose of information processing and eventually intelligence \cite{kagan_two_2025}. Broadly, some approaches focus on physiological relevance in an approach termed "Organoid Intelligence (OI)" (e.g.\cite{alam_el_din_human_2025, smirnova_organoid_2023}) while others aim to use highly structured networks with distinct properties, which has been termed "Bioengineered Intelligence (BI)" (e.g. \cite{sumi_biological_2023, kagan_two_2025}). Given that closed-loop input and output with even simple neural cultures can lead to dynamic network-wide effects \cite{habibollahi_critical_2023}, there will be a growing focus on predictive metrics which can identify which network structures inherently possess the ability to modulate input information in diverse ways. Even at the simplest level, the terminology to be able to easily communicate the information-processing order that a given neural culture is undertaking will be useful as a metric to explain the complexity of whatever task or process is being tested.

\subsection{Substrate Independent Comparisons}

Meaningfully comparing different system architectures is difficult, especially when substrates are fundamentally different but the surface level outcome might appear similar \cite{voges_decomposing_2024, dale_substrate-independent_2019}. For example, when comparing biological learning to machine-learning methods, comparisons can always be made on performance \cite{khajehnejad_dynamic_2025}, but the architecture and internal dynamics of the systems differ so greatly that further comparisons in how information is handled at each time step are typically limited. The benefit of using substrate-independent metrics is that it allows comparisons in systems that may dramatically differ in most respects. However, by identifying the consistent treatment of incoming information within the system, more relevant comparisons can be made. Similar substrate-independent approaches in the past have provided other metrics that are useful in this way, such as the information flow from the external environment to the internal system \cite{kolchinsky_semantic_2018}. A key difference in the approach we propose is that it focuses not on what information is maintained by the system, but on what is transformed through the information-processing architecture of the system. The proposed framework of information-processing orders also differs to the similarly named approach that aims to quantify the information-processing capacity (IPC) of dynamic systems \cite{dambre_information_2012, schulte_to_brinke_refined_2023}. As a metric, IPC enables investigations of the dynamical systems in terms of the polynomial functions that can be computed and the memory required for this task \cite{schulte_to_brinke_refined_2023}. However, it does not explore the actual transformation of the information by the system outside of these constraints, nor establish a framework for the different orders of information processing and how they may relate to other system properties. Likewise, a range of other information dynamic metrics have also focused on low-level processes such as storage, transfer, and modification of information by a system \cite{voges_decomposing_2024} or other metrics such as semantic information \cite{kolchinsky_semantic_2018}. The framework we propose here is not intended to stand-alone, but rather be augmented by the use of these existing substrate independent metrics. By combining these different substrate-neutral information measures and approaches, one can identify common principles and differences in how information is handled across the range of possible systems that could be explored. In effect, these metrics permit a consistent treatment of incoming information by a system’s architecture, making it possible to draw more meaningful comparisons between systems that otherwise have little in common.

\subsection{Applicability to Neuroscience}
The exploration of how a system processes information is particularly focused in the areas of neuroscience. Neurocomputational metrics for information-processing capacity have previously been proposed which aim to provide a quantification of the complexity of a system in terms of the functions that the system can compute \cite{schulte_to_brinke_refined_2023}. Here the information-processing order framework does not focus on the eventual functional outcome of a process, but the system-level information transformation which occurs. This complimentary perspective is likely critical, as the ability for a system to actively transform input has been identified as required for more complex learning tasks \cite{miconi_neural_2025}. There is ample evidence that biological neural systems possess all classes of information processing that are proposed in this paper, with an interaction between different types of systems leading to yet more complex behaviours in a controlled fashion \cite{toyoizumi_modeling_2014, pariz_high_2018, bastos_visual_2015}. Finally, this system-based metric would compliment other areas in clinical and pre-clinical neuroscience research. Take for example the Perturbational Complexity Index (PCI), which aims to assess the normalized compressibility of an input signal throughout the brain \cite{sinitsyn_detecting_2020}. It has been proposed that increased PCI scores predict consciousness (in the medical sense) in otherwise comatose patients \cite{sinitsyn_detecting_2020}. In theory any class of system presented in this framework may alter the compressibility of inputted signal - even if only by outputting a randomised response throughout the system once a given threshold of stimulus is presented. However, it might be possible to augment the stimulus to explore whether a Class I, Class II, or Class III system dynamic is present in the brain of these patients, and further understand the function that may be present in otherwise minimally responsive patients.

% \begin{itemize}
%     \item \textbf{Input-vs-Output, loop area hysteresis.} Definition: Signed area of the input--output curve. {\color{blue}[Add mathematical definition]}, to get memory of each system at different orders of complexity
%     \item \textbf{Input-vs-Output slope drift}. Definition: 
% \end{itemize}

% %%--------------------------------------------%%
% \section{Figures \& Tables}
% %%--------------------------------------------%%
% The output for figure is:\vspace*{-7pt}

% \begin{figure}[!h]
% %\centering\includegraphics[width=2.5in]{xxxxxx.eps}
% %%% where xxxxxx name represents "figurename.eps"
% \caption{Insert figure caption here}
% \label{fig_sim}
% \end{figure}

% \vspace*{-5pt}

% \noindent The output for table is:\vspace*{-7pt}

% \begin{table}[!h]
% \caption{An Example of a Table}%%%Table caption goes here
% \label{table_example}
% \begin{tabular}{llll}%%%The number of columns has to be defined here
% \hline
% date &Dutch policy &date &European policy \\
% \hline
% 1988 &Memorandum Prevention &1985 &European Directive (85/339) \\
% 1991--1997 &{\bf Packaging Covenant I} & & \\
% 1994 &Law Environmental Management &1994 &European Directive (94/62) \\
% 1997 &Agreement Packaging and Packaging Waste & & \\\hline
% \end{tabular}
% \vspace*{-4pt}
% \end{table}%%%End of the table

\section{Conclusion}
By proposing the addition of a necessary condition that a system must display to be qualified as having agency, this work aims to resolve disagreements in the field and provide a useful framework in which to discuss the dynamics of these systems. By proposing a bottom-up system dynamic approach that can be integrated with existing top-down capacity-based definitions, consistent and useful definitions of agency can be formulated. The information-processing order framework has direct implications in discussing the ethical considerations for if a system qualifies for agency, and will allow more nuanced discussions on what moral considerations may therefore apply if so. However, it should be acknowledged that even if such systems do display high levels of agency this is not equivalent to a moral status. Here, we suggest that it is only with the advent of conscious agency that a system would have moral status, and attendant rights. Yet, by establishing testable and falsifiable frameworks that allow terms such as agency to be explored in greater depth, progress towards identifying metrics that indicate these other morally relevant traits can also progress. Additionally, the framework may help in predicting the functional potential of systems while also allowing substrate-independent comparisons of information-processing dynamics. More broadly, this framework is also applicable to neuroscience in describing what information dynamics neural systems and their subsystems are undertaking when processing information. Finally, for work seeking to use in vitro neural cultures for information processing and intelligence, the proposed framework offers a way to assess what information-processing order these biological systems are undertaking when responding to structured information environments. This will clarify which tasks involve agentic steps compared to those that are less complex. Ultimately, with the growing interest, progress, and investment in developing autonomous and intelligent systems, the proposed framework offers a principled and testable pathway to resolve both uncertainty and existing disagreements related to terms such as agency, along with clarifying the internal dynamics these systems may display.

\subsection{Acknowledgments}{The authors thank and acknowledge Andy C. Kitchen for insightful discussions and suggestions and Dr Nicole Kerlero de Rosbo for proofreading. This research is supported by NUSMed and ODPRT (NUHSRO/2024/035/Startup/04) for the project “Experimental Philosophical Bioethics and Relational Moral Psychology” with BDE as PI. This research is supported by the National Research Foundation, Singapore under its AI Singapore Programme (AISG Award No: AISG3-GV-2023-012). This research project is supported by National University of Singapore under the NUS Start-Up grant; (NUHSRO/2022/078/Startup/13). This work was supported by the Wellcome Trust [Grant number: 226801] for Discovery Research Platform for Transformative Inclusivity in Ethics and Humanities Research (ANTITHESES). For the purpose of open access, the author has applied a CC BY public copyright licence to any Author Accepted Manuscript version arising from this submission. This project is supported by the National Science Foundation grant entitled “EFRI BEGIN OI: Bringing Reward to Embodied Organoid Intelligence” (FAIN \#: 2515214) to JLB as Co-PI.}

%%% Appendix %%%

\appendix
\section{Appendix}

\subsection{ Rolling regression}
\label{sec:app_roll_regression}

To compute the local gain $m(t)$ and local bias $b(t)$, we used the sliding-window linear regression approach described section~\ref{sec:rollfit}. Within each time window, we regressed the output onto the input, providing time-resolved estimates of how strongly the system responds to the stimulus ($m(t)$), and around which baseline level ($b(t)$).

A key consideration is the choice of window length. If the window is too short, the input within that interval may not vary sufficiently, particularly during low-frequency segments, leading to unstable or noisy estimates. If the window is too long, genuine temporal evolution in the transformation may be averaged out, and adaptive effects may be artificially smoothed away.

Because the driving signal is a slow-varying sinusoidal sweep spanning nearly two orders of magnitude in frequency ($0.1 \,\si{Hz}\rightarrow 10 \, \si{Hz}$), we tied the window size to the slowest frequency present. The lowest-frequency segment is $f_0 = 0.1\,\si{Hz}$, corresponding to a period of $T_0 = 10\,\si{s}$. A 4-$\si{s}$ window ($ W \approx 0.4 \, T_0$) ensures two important conditions:
\begin{itemize}
    \item In the lowest-frequency regime, each window captures a meaningful fraction of a cycle, giving enough variation for a well conditioned linear fit.
    \item For the mid and high-frequency regimes, each window spans multiple cycles, reducing variance in the resulting estimates of $m(t)$ and $b(t)$.
\end{itemize}
The window was advanced in steps of approximately $S= 1\,\si{s}$ (about $0.1 \ T_0$). This choice provides dense temporal sampling with substantial overlap between successive windows, enabling slow drifts in the estimated gain and bias to be tracked continuously without introducing unnecessary redundancy.
\begin{figure}[!h]
\centering
\begin{subfigure}[t]{0.49\linewidth}
    \centering
    \includegraphics[width=\linewidth]{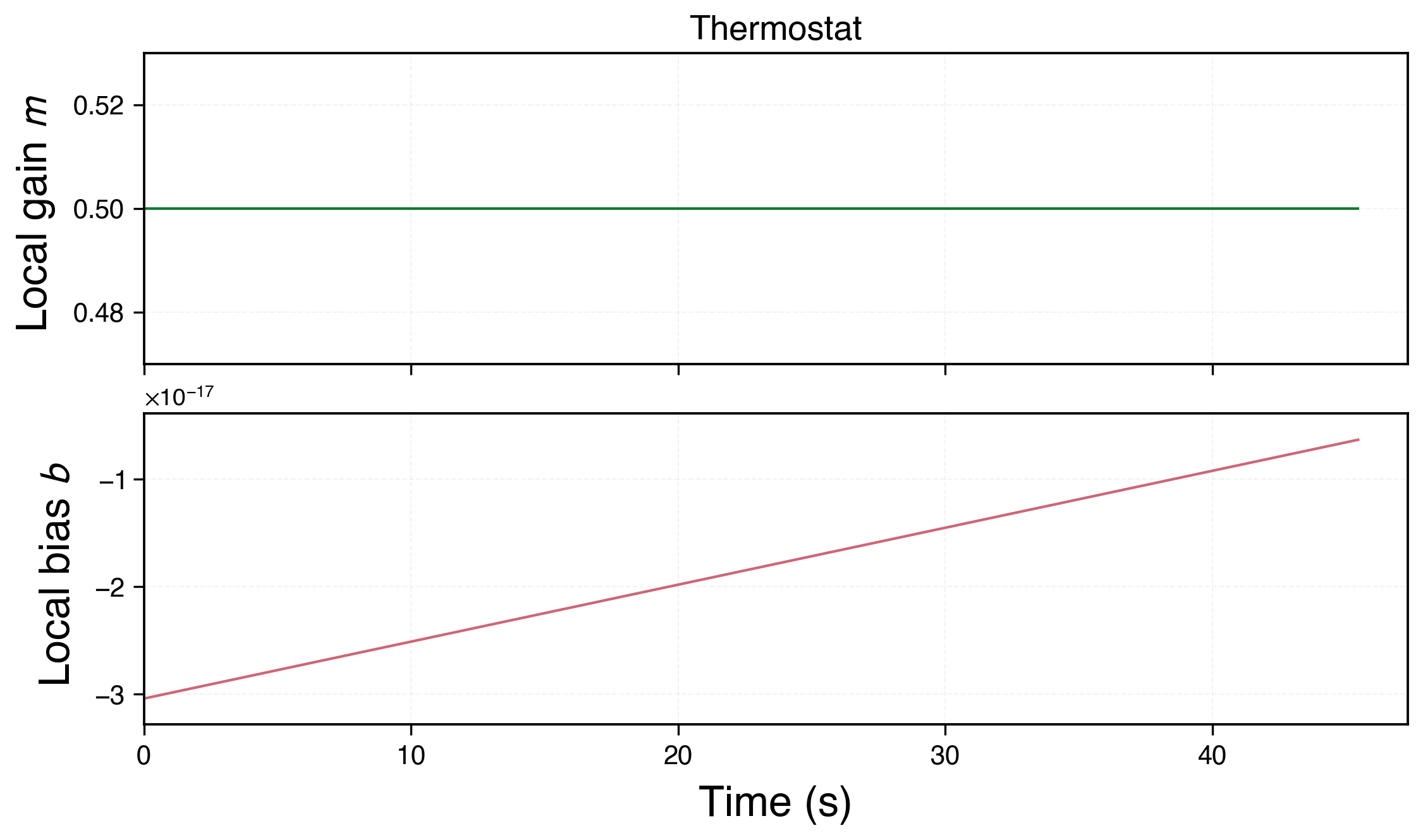}
    \caption{Thermostat}
    \label{fig:rolling-fit-thermo}
\end{subfigure}
\hfill
\begin{subfigure}[t]{0.48\linewidth}
    \centering
    \includegraphics[width=\linewidth]{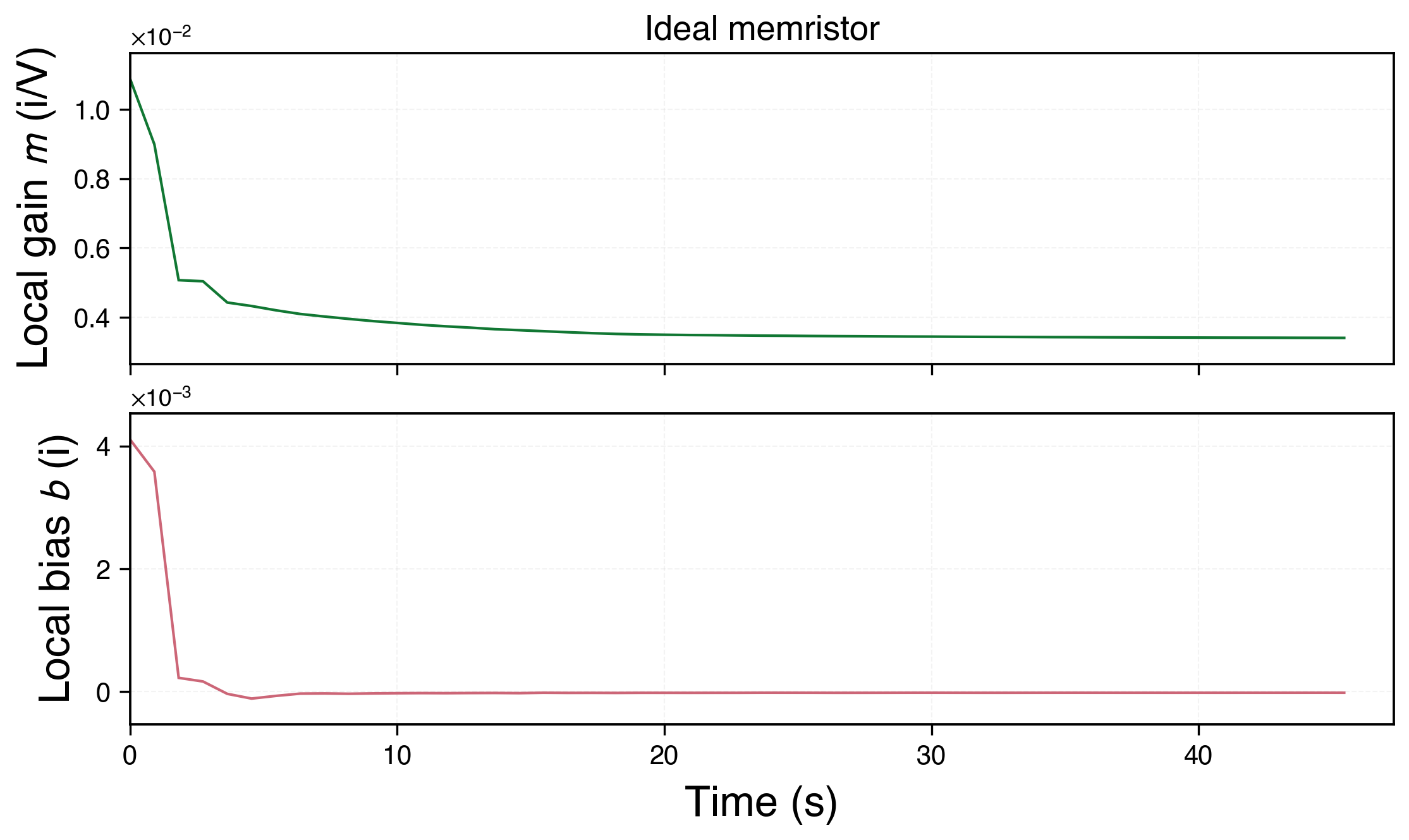}
    \caption{Ideal memristor}
    \label{fig:rolling-fit-nonadapt}
\end{subfigure}
\vspace{0.8em}
\begin{subfigure}[t]{0.50\linewidth}
    \centering
    \includegraphics[width=\linewidth]{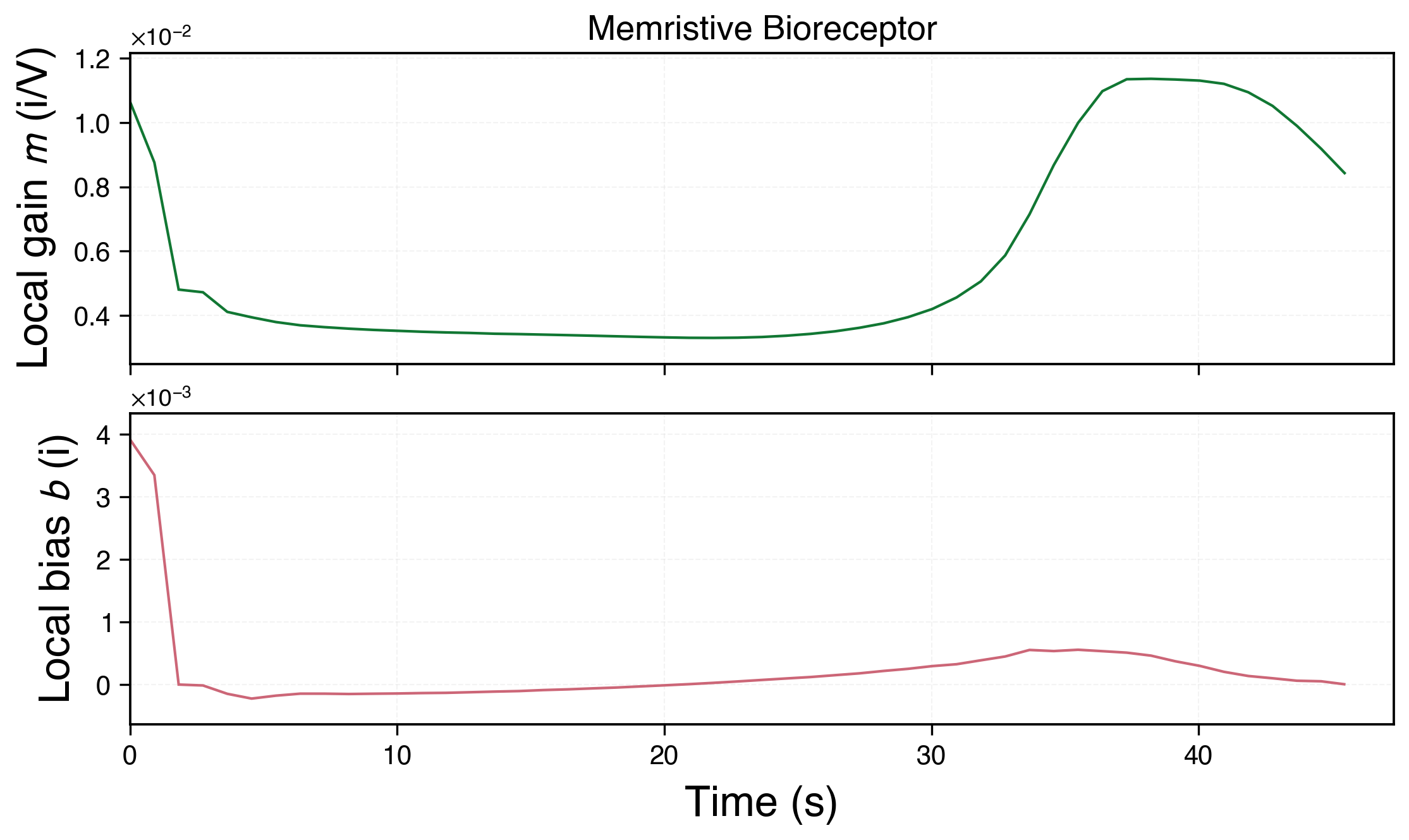}
    \caption{Memristive bioreceptor}
    \label{fig:rolling-fit-adaptive}
\end{subfigure}
\caption{Rolling gain and bias for all systems: the thermostat is constant, the ideal memristor slowly relaxes, and only the MBR shows genuine time-dependent change. }
\label{fig:rolling-fit-all}
\end{figure}

The same window and step parameters were used for all systems, (thermostat, ideal memristor, and memristive bioreceptor), ensuring same analysis settings for all systems. The complete set of rolling gain and bias trajectories for the three systems is presented in Fig.~\ref{fig:rolling-fit-all}, illustrating the basis for the comparisons discussed above. The thermostat exhibits a constant local gain and bias throughout; the ideal memristor shows slow state-dependent relaxation but no adaptation; and the memristor bioreceptor displays pronounced temporal evolution in both parameters, reflecting its adaptive internal dynamics.
%

% \begin{figure}[h!]
%     \centering
%     \includegraphics[width=0.8\linewidth]{Figures/thermo_sv_gain_bias_stacked.png}
%     \caption{}
%     \label{fig:rolling-fit-thermo}
% \end{figure}

% \begin{figure}[h!]
%     \centering
%     \includegraphics[width=0.8\linewidth]{Figures/ideal_mem_sv_gain_bias_stacked.png}
%     \caption{}
%     \label{fig:rolling-fit-nonadapt}
% \end{figure}
% %
% \begin{figure}[h!]
%     \centering
%     \includegraphics[width=0.8\linewidth]{Figures/adapt_mem_sv_gain_bias_stacked.png}
%     \caption{Enter Caption}
%     \label{fig:rolling-fit-adaptive}
% \end{figure}

\subsection{Zero-crossing lag}
\label{app:zerocrossing}

\begin{figure}[!h]
    \centering
        \begin{subfigure}{0.47\linewidth}
        \includegraphics[width = \linewidth]{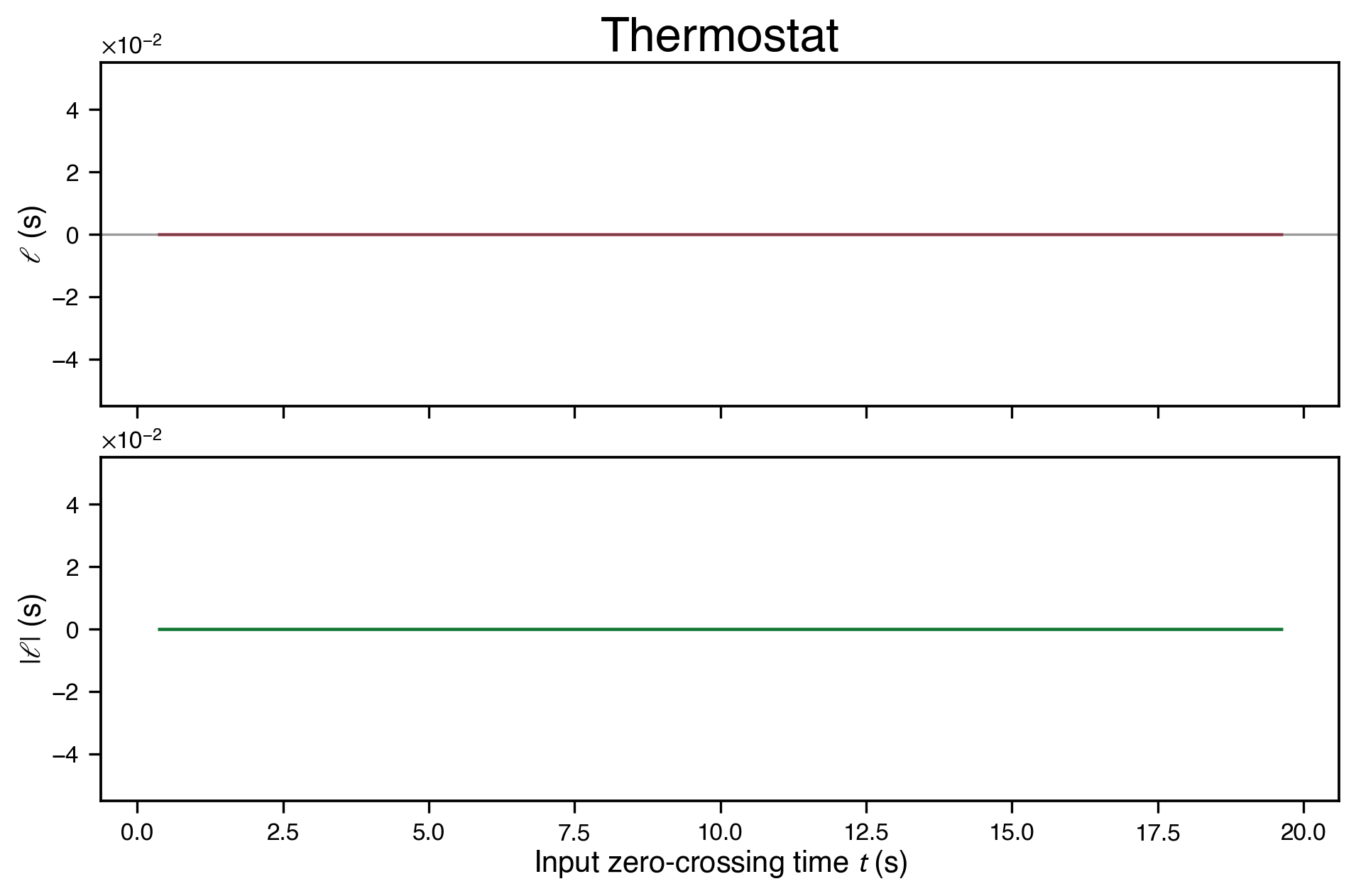}
        \caption{Thermostat}
        \label{fig:zerocross-thermo}
    \end{subfigure}
    \begin{subfigure}{0.47\linewidth}
        \includegraphics[width= \linewidth]{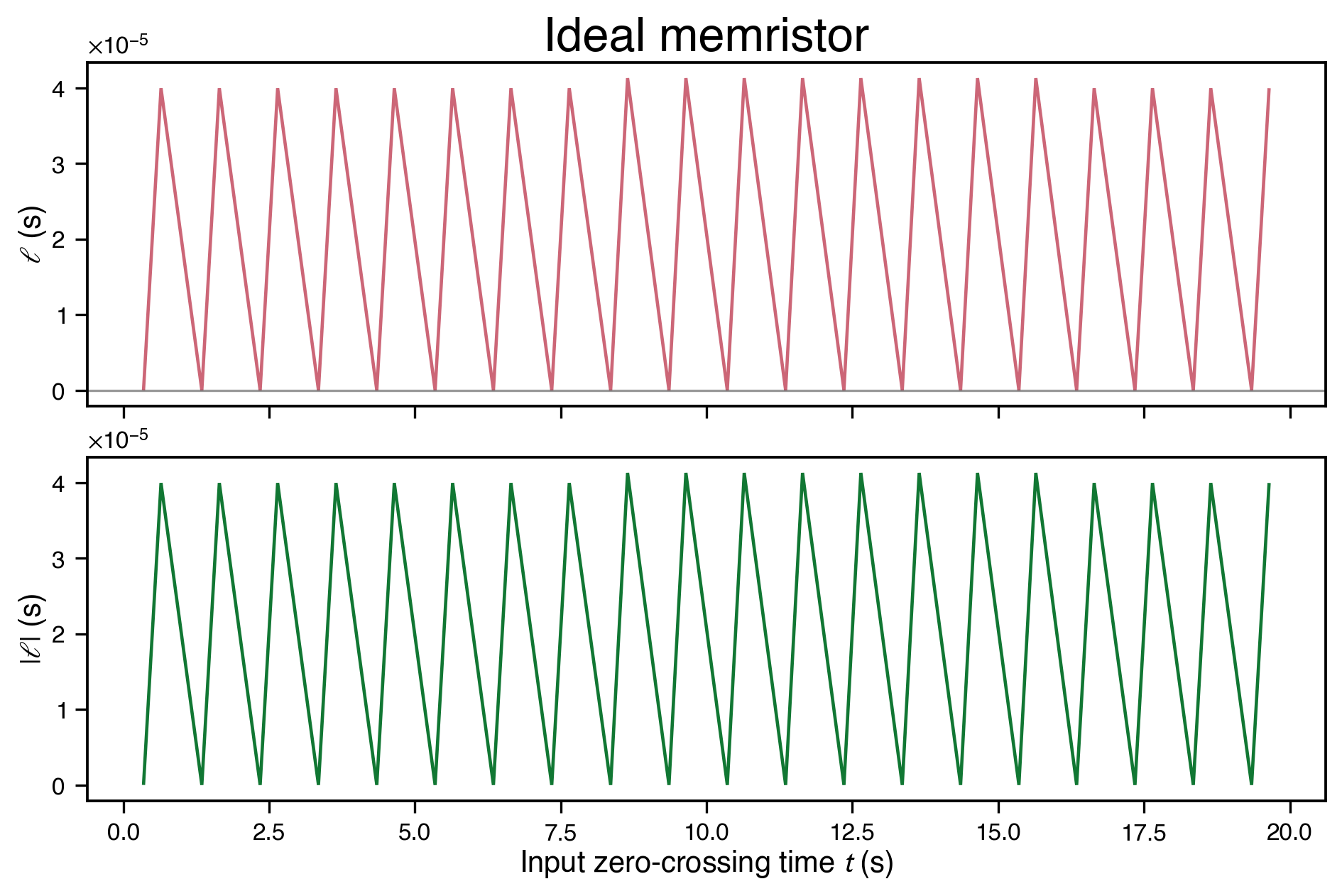}
        \caption{Ideal memristor}
        \label{fig:zerocross-ideal}
    \end{subfigure}
    \caption{Zero-crossing lag for thermostat and ideal memristor. The thermostat exhibits identically zero lag at all crossings, while the ideal memristor shows a small but fixed offset whose magnitude remains constant over time.}
    \label{fig:zc_appendix}
\end{figure}
Figures~\ref{fig:zc_appendix} shows the zero-crossing lag for the thermostat and ideal memristor driven by a bipolar square-wave input. As expected, the thermostat displays identically zero lag at all crossing, while the ideal memristor shows a small but fixed time offset whose magnitude remains constant over time. Neither system displays the progressive drift in $|\ell^{(k)}|$ characteristic of the memristive bioreceptor. Their timing alignment remains stationary throughout.

%\bibliographystyle{unsrtnat}
%\bibliography{references1}  %%% Uncomment this line and comment out the ``thebibliography'' section below to use the external .bib file (using bibtex) .

%Uncomment this section and comment out the 

%\bibliography{references} line above to use inline references.




% \end{thebibliography}

\end{document}